# Metareciclagem no ensino de Física na Educação Básica no contexto dos Objetivos de Desenvolvimento Sustentável da ONU a partir de pesquisa tecnológica

Metarecycling in Physics Education in Basic Education within the Context of the UN Sustainable Development Goals through Technological Research

**James A. Souza[1]**
Departamento de Física, Universidade Federal de São Carlos – UFSCar

**Michel Corci Batista**
Departamento Acadêmico de Física, Universidade Tecnológica Federal do Paraná – UTFPR

**Marcello Ferreira**
Instituto e Centro Internacional de Física, Universidade de Brasília – UnB

**RESUMO**

O rápido avanço das tecnologias digitais no primeiro quarto do Século XXI introduziu transformações significativas em várias áreas, como comunicação, saúde e educação, mas gerou aumento no uso e descarte de dispositivos eletrônicos, resultando em desafios ambientais relacionados aos Resíduos de Equipamentos Elétricos e Eletrônicos (REEE) – também conhecido como *e-lixo*. Isso é observado nas escolas, espaço em que a inserção e renovação de equipamentos têm se tornado demanda para o desenvolvimento e a aplicação de novas estratégias de ensino. A partir de uma pesquisa tecnológica, apresentamos como estudantes de uma escola pública do interior de São Paulo conduziram processos de reaproveitamento do *e-lixo*, utilizando os preceitos da metareciclagem e conhecimentos de Física para construir uma bateria portátil (*powerbank*) e um carregador de *smartphones* alimentado por um dínamo acoplado a uma bicicleta. A apropriação das relações entre ciência, tecnologia e aspectos sociais foi conduzida a partir da validação dos carregadores por meio de testes de caracterização do tempo de carga oferecido pelo *powerbank* e pelo dispositivo instalado na bicicleta em condições de passeio. As ações educativas na comunidade, envolvendo noções de sustentabilidade, de energia limpa e os benefícios à saúde pela prática de exercícios físicos tiveram como referência os Objetivos de Desenvolvimento Sustentável (ODS) da ONU.

**Palavras-chave:** Energia limpa. Ensino de Física. Metareciclagem. Tecnologia. Sustentabilidade.

**ABSTRACT**

The rapid advancement of digital technologies in the first quarter of the 21st century has introduced significant transformations in various fields, such as communication, healthcare, and education. However, it has also led to an increase in the use and disposal of electronic devices, resulting in environmental challenges related to Waste Electrical and Electronic Equipment (WEEE) – also known as e-waste. This phenomenon is observed in schools, where the integration and renewal of equipment have become essential for the development and implementation of new teaching strategies. Based on a technological research project, we present how students from a public school in São Paulo's countryside conducted e-waste reuse processes, applying the principles of metarecycling and physics knowledge to build a portable battery (power bank) and a smartphone charger powered by a dynamo attached to a bicycle. The appropriation of the relationships between science, technology, and social aspects was facilitated by validating the chargers through characterization tests of the charging time provided by the power bank and the bicycle-installed device under riding conditions. Educational actions within the community, involving concepts of sustainability, clean energy, and health benefits through physical exercise, were guided by the United Nations Sustainable Development Goals (SDGs).

---

[1] Endereço de contato: jasouza@ufscar.br.

## 1. Introdução

A indústria eletrônica é a que apresenta o crescimento mais rápido entre os setores manufatureiros em todo o mundo. Ela é responsável pela fabricação, montagem e manutenção de produtos eletrônicos, os quais possuem diversidas de aplicações, abrangendo componentes discretos, como circuitos integrados; eletrônicos de consumo, como TVs, *smartphones* e computadores pessoais; equipamentos médicos, como monitores de frequência cardíaca, máquinas de diálise e aparelhos para formação de imagens; equipamentos industriais, como robôs; equipamentos de comunicação e rede, como roteadores e painéis de distribuição, entre muitos outros. Isto é, o crescimento vertiginoso da indústria eletrônica nas últimas duas décadas, sobretudo, é consequência de profundas mudanças socias e culturais marcadas pela cibercultura, pelas inteligências coletivas e pelas maneiras de comunicação em redes (Lévy, 2003; 2010), bem como pela dependência de outras indústrias, como a automotiva, de avaliação, de defesa, de telecomunicações, de entretenimento e de saúde.

A consequência do seu crescimento, combinado com a rápida obsolescência dos produtos, os avanços tecnológicos e o aumento do consumo constituem um novo desafio ambiental devido à grande geração de Resíduos de Equipamentos Elétricos e Eletrônicos – REEE (Baldé *et al.*, 2017; Ilanknoon *et al.*, 2018; Forti *et al.*, 2020; Aguiar *et al.*, 2021). O volume de REEE, também conhecido como *e-lixo*, gerado em 2019 foi de 53,6 milhões de toneladas. Esse aumento é superior a 20% desde 2014 e equivale, em média, a geração de 7,3 quilogramas de *e-lixo* em 2019 por cada pessoa do planeta. Destes, 1,7 quilograma *per capita*, isto é, apenas 23,3%, foram descartados de maneira ambientalmente sustentável (UN DESA, 2021).

A produção de *e-lixo* é uma preocupação crescente para a maioria dos países, concentrando-se principalmente em regiões com maior crescimento econômico e urbano. O Brasil, por exemplo, está entre os maiores produtores mundiais de equipamentos eletrônicos e estima-se que apenas 3% do *e-lixo* decorrente seja descartado de maneira adequada (Baldé *et al.*, 2017; Franz, 2021). A presença de materiais como metais preciosos e substâncias tóxicas no *e-lixo*, por seu turno, apresenta oportunidades de reciclagem, mas riscos ambientais graves, com impactos diretos no solo, na água e na saúde humana (Lu *et al*., 2014).

No Brasil, a metareciclagem, uma prática que consiste no reaproveitamento de componentes eletrônicos descartados para a criação de novos dispositivos, surge como uma solução inovadora para mitigar os impactos ambientais e promover a sustentabilidade. No contexto educacional, esta pode ser utilizada como uma ferramenta pedagógica poderosa, proporcionando aos estudantes a oportunidade de aplicar conceitos de ciências e tecnologias de maneira prática e interdisciplinar para o desenvolvimento de protótipos e soluções tecnológicas, alinhando o ensino à sustentabilidade e a Objetivos de Desenvolvimento

Sustentável (ODS) da ONU.

Nesse contexto, a pesquisa tecnológica relatada neste artigo trata da metareciclagem como uma prática sustentável e educativa no âmbito da Educação Básica, especificamente com conceitos da área de Física. Ela foi desenvolvida com estudantes de uma escola pública do interior do estado de São Paulo. A partir de dispositivos eletrônicos descartados, eles construíram protótipos, como carregadores de *smartphones* e *powerbanks*, desenvolvendo soluções práticas para a geração de energia limpa e sustentável. Essa prática foi alinhada com os ODS 3 (vida saudável), ODS 4 (educação de qualidade) e ODS 7 (energia acessível e limpa), contribuindo para a conscientização ambiental, o uso responsável de tecnologias e a promoção da sustentabilidade.

Adicionalmente, é explorado como essa experiência pode ser incorporada ao ensino de Física, promovendo aprendizagem interdisciplinar por meio da investigação científica, da prática de soluções tecnológicas e do desenvolvimento de competências no uso de tecnologias digitais no processo educativo. Nosso objetivo é evidenciar o potencial da metareciclagem não apenas como uma alternativa ambientalmente sustentável, mas como uma abordagem pedagógica inovadora no ensino de ciências.

## 2. A Metareciclagem no Ensino de Ciência e Tecnologia

A crescente geração de *e-lixo*, as baixas taxas de coleta e reciclagem e o seu descarte inadequado vêm colocando em risco o meio ambiente, a diversidade biológica, a produção agropecuária e a saúde humana. A previsão, para 2030, é que cada pessoa no planeta produza até 9,0 quilogramas de *e-lixo*, o que equivaleria a 74,4 milhões de toneladas. Em contrapartida, na última década, a taxa anual *per capita* de crescimento da reciclagem desses resíduos foi de apenas 0,05 quilograma (UN DESA, 2021). Essas estimativas sugerem a urgente necessidade de estabelecer estratégias para minimizar a produção, aperfeiçoar a legislação e a gestão do *e-lixo* no planeta e incorporar novas práticas pedagógicas no ensino de ciência e tecnologia, a fim de sensibilizar os estudantes para a importância da sustentabilidade e capacitá-los a desenvolver soluções inovadoras para problemas ambientais, como a gestão do lixo eletrônico. No campo educacional, a gestão e a valorização do *e-lixo* podem ser abordadas de maneira interdisciplinar[2], por exemplo, nas disciplinas de Física, Química, Biologia, Matemática, Ciências Ambientais, Ética e Relações Internacionais, tendo potencial significativo para beneficiar a aprendizagem e a conscientização crítica dos estudantes. As tecnologias em desenvolvimento, como a Internet das Coisas, as *blockchain*, a nanotecnologia e conceitos como cidade inteligente, computação e economia verde, indústria ecológica, cidade sustentável etc., podem ser todas abordadas de maneira combinada nas escolas para que essas ideias e práticas sejam valorizadas e proliferadas na vida dos estudantes.

Uma alternativa de reciclagem do *e-lixo* é o desmonte e o reaproveitamento de dispositivos e respectivos componentes descartados. No Brasil, essa prática é conhecida como metareciclagem e vem progressivamente ganhando espaço devido ao seu potencial em relação aos cuidados com o meio ambiente em prol de comunidades sustentáveis e de transformação social (Neto, 2022). Com a metareciclagem é possível desenvolver, a partir de tecnologias e sistemas eletrônicos descartados, protótipos de dispositivos com o reaproveitamento de componentes eletrônicos, carcaças e peças diversas.

Na Educação Básica (EB), a metareciclagem pode ser extremamente vantajosa para o ensino de ciências e tecnologiasem abordagens como a da aprendizagem *maker*, caracterizada pelo desenvolvimento de projetos para a solução de problemas concretos, envolvendo interação, dialogia, investigação, interdisciplinaridade, formulação de hipóteses, elaboração de esquemas para coleta e análise de dados, interpretação, extrapolação e comunicação científica de resultados a questões de interesse e relevância. O estabelecimento da cultura e de espaços *maker* nas escolas pode favorecer o protagonismo do estudante para a criação e o surgimento de ideias em um ambiente colaborativo, exercendo a sua criatividade e autonomia (Halverson; Sheridan, 2014).

---

[2] Para as finalidades deste trabalho, tomaremos uma noção limitada de interdisciplinaridade, porém, operacional e eficiente, como a epistemologia aplicável à solução de problemas complexos a partir da integração discursiva, prolongada e articulada, de campos dos saberes (e respectivas epistemologias, teorias e metodologias), produzindo efeitos na compreensão de um fenômeno, sem, com isso, modificar as disciplinas-base (Sommerman, 2015; Silva Filho & Ferreira, 2018; Gulis *et al*., 2021; Ferreira *et al*., 2022).

O uso da internet na EB, impulsionada pelo distanciamento social durante a pandemia de Covid-19 nos anos de 2020 e 2021, induziu aumento significativo no uso de *smartphones*, *tablets*, computadores pessoais e *laptops* por parte dos estudantes. Para cumprir atividades pedagógicas *on-line*, síncronas (Oliveira; Mill; Ferreira, 2023) ou assíncronas (Mill; Oliveira; Ferreira, 2022), equipamentos eletrônicos foram introduzidos para a realização de apresentações multimídia, a produção de vídeos e o funcionamento de aplicativos e plataformas digitais. Muitas dessas demandas agregaram-se definitivamente aos processos de ensino-aprendizagem, incorrendo no aumento de dispositivos eletrônicos nas escolas e na consequente ampliação do seu descarte, dada a obsolescência programada[3] desse tipo de recurso.

Diante desse cenário, estudantes da Escola Estadual Professor Roque Conceição Martins, localizada na cidade de Sorocaba, no estado de São Paulo, idealizaram utilizar partes dos dispositivos descartados para construir soluções sustentáveis para carregamento portátil de *smartphones* e *tablets*. Precocemente acessíveis a crianças e adolescentes, eles oferecem crescente variedades de aplicativos para estudos e pesquisas, conexão à internet, comunicação, vinculação a mídias socias e registro de fotos e vídeos, aumentando a demanda por fontes energéticas para carregamento. Além de explorar a vantagem da mobilidade com estes carregadores, tiveram como meta desenvolver dispositivos que reduzissem riscos ambientais e ecológicos, utilizando energia limpa de maneira sustentável.

Para isso, utilizaram como material de orientação os Objetivos do Desenvolvimento Sustentável (ODS) estabelecidos pela Organização das Nações Unidas (ONU) para proteger o meio ambiente e o clima e garantir qualidade devida e educação de qualidade para todos (UN DESA, 2021). Diante desse desafio, pojetaram inicialmente um dispositivo que favorecesse a concretização do ODS 3 – "assegurar uma vida saudável e promover o bem-estar para todos, em todas as idades" (UN DESA, 2021, p. 30). Em outra linha, na tentativa de estimular a prática de exercícios físicosna escola, construíram um dispositivo carregador de *smartphones* e *tablets* utilizando um dínamo acoplado a uma bicicleta estacionária, cujo princípio de funcionamento é a conversão da energia mecânica, proveniente das pedaladas, em elétrica. Apesar de cumprir de maneira eficiente com o seu propósito, esse dispositivo não atendia à necessidade de mobilidade da maioria dos estudantes. A partir da proposta de metareciclagem, foi construído outro aparelho carregador, o *powerbank*, bateria externa dispositivo simples e compacta. Além de poder ser levado a qualquer lugar, permite carregar aparelhos em situações em que não há disponíveis fontes usuais (como tomadas elétricas).

A construção dos dois dispositivos promoveu, em um espaço *maker*, a oportunidade de aprendizagem de diversos conceitos científicos e tecnológicos, buscando alinhamento ao ODS

---

[3] Por limitações de escopo, não dedicaremos tratamento profundo a essa questão. Dela, o relevante é ter em conta que, no modo de produção capitalista, os bens de consumo são produzidos com expectativa prévia – e breve – de inutilização, para consequente substituição. Isso pode se dar por meio da adoção de materiais de baixa qualidade ou pouco duráveis ou pela configuração de vínculos de atualização obrigatória (como é o caso das diversas versões dos *softwares* que acompanham tais recursos e que, a certa altura, passam a ser incompatíveis com a versão do dispositivo adquirido, levando à redução de funcionalidades ou mesmo à inviabilização do uso). A obsolescência programada, portanto, é um aspecto que interage diretamente com as noções de *e-lixo*.

4 – "assegurar a educação inclusiva, equitativa e de qualidade, e promover oportunidades de aprendizagem ao longo da vida para todos" (UN DESA, 2021, p. 34).

Nessa perspectiva, foi possível notar a valorização de disciplinas científicas no terceiro ano do Ensino Médio, em especial a de Física, pois além de exigir dos estudantes o conhecimento de conceitos, essa aplicação ocorreu de maneira interdisciplinar, mobilizando habilidades e competências inerentes ao fazer científico como conhecimento, método e cultura, tais como: desenvolver a capacidade de investigação por meio da observação, classificação, organização e sistematização; compreender o conceito de medir, analisando diferentes escalas, unidades e ordens de grandeza; formular e testar hipóteses; propor modelos explicativos e representativos para os sistemas tecnológicos estudados; identificar parâmetros relevantes para o funcionamento e a caracterização dos dispositivos construídos; quantificar grandezas e relacioná-las com fatores de qualidade para conferir maior eficiência aos dispositivos, investigar situações-problema, propor soluções e realizar comunicação científica.

A proposta apresentada, ademais, visa ao acesso à energia de maneira confiável, sustentável e moderna para todos, conforme o ODS 7 – "garantir acesso à energia barata, confiável, sustentável e renovável para todos", além de ser promissora para conceber cidades e comunidades sustentáveis, em pleno acordo com o ODS 11 – "tornar as cidades e os assentamentos humanos inclusivos, seguros, resilientes e sustentáveis" (UN DESA, 2021, p. 40-48). As competências e habilidades desenvolvidas a partir do estudo de conceitos específicos das ciências da natureza e suas tecnologias, aliadas à compreensão dos ODS, podem ajudar os estudantes a modificarem posturas frente à gestão do lixo eletrônico produzido na escola e na comunidade.

## 3. Métodos e Materiais

A pesquisa aqui relatada é classificada como tecnológica, pois emprega meios de pesquisa científica (básica ou aplicada) com o propósito de construir um artefato cuja composição e utilidade dela derive (Freitas Júnior & Sousa, 2018; Ferreira *et al.*, 2022). O desenvolvimento de dispositivos a partir do *e-lixo* coaduna com a perspectiva de tecnologia de Bunge (1985, p. 231): "[...] campo do conhecimento relativo ao projeto de artefatos e ao planejamento de sua realização, operação, ajuste, manutenção e monitoramento, à luz do conhecimento científico". Como pesquisa tecnológica, deve prever as etapas de: 1) identificação do problema, motivação e sensibilização; 2) definição dos objetivos da tecnologia proposta; 3) *design* e desenvolvimento do artefato, envolvendo o mapeamento de construção e a modelagem de evidências, o modelo e/ou os mapas de modelos das tarefas, a projeção do modelo de implementação e a calibragem por meio de controle de qualidade e avaliação; 4) demonstração, isto é, a implementação do artefato, em ambiente controlado, mais próximo o possível do real, realizando testes de desempenho, conferência de performance e avaliação de alcance dos objetivos; e, por fim, 5) avaliação, etapa em que o desenvolvedor e os usuários buscam analisar o artefato a partir da sua funcionalidade adotando indicadores como usabilidade, adequação de requisitos, precisão, viabilidade, completude, sustentabilidade e segurança (Freitas Júnior & Sousa, 2018).

Nessa perspectiva, para a montagem dos dispositivos, os estudantes testaram diferentes materiais e configurações, com o intuito de fazer com que o produto tivesse baixo custo e um *design* ergonômico, de maneira que a sua utilização fosse fácil e trouxesse conforto para o usuário durante a utilização. As atividades foram desenvolvidas principalmente no contraturno das aulas e nas disciplinas "Cálculos na Otimização de Resultados" e "Luz e Tecnologia", que compõem os itinerários formativos do currículo da escola no âmbito do Novo Ensino Médio. Os dispositivos foram construídos manualmente, utilizando ferros de solda para a montagem dos circuitos elétricos, pedaços de madeira para a construção da base estacionária da bicicleta e uma impressora 3D para a confecção de uma caixa para a inserção do circuito elétrico e das baterias do *powerbank*, buscando visual atraente. Todos os materiais foram obtidos por meio de doações, fornecidos aos estudantes como sucata, sendo restaurados e reaproveitados seguindo os preceitos da metareciclagem.

Para implementar a metareciclagem no ensino de Física, foi desenvolvido um plano de ação detalhado que se desdobra em várias etapas. Inicialmente, os estudantes participaram de uma sessão introdutória acerca dos conceitos de sustentabilidade, utilizando materiais audiovisuais para contextualizar a importância do reaproveitamento de componentes eletrônicos. Em seguida, realizou-se prática de 2 semanas, com encontros de 2 horas cada, em que dispositivos eletrônicos descartados, como computadores e celulares, foram desmontados para identificar e separar componentes reutilizáveis. Além de explorar e documentar suas descobertas, foram orientados a projetar e construir dispositivos, como carregadores ou pequenas fontes de energia, utilizando os materiais recuperados. Para isso, foram utilizadas ferramentas básicas de eletrônica, como ferros de solda e multímetros, além de materiais adicionais como fios e conectores. Esta fase do projeto teve uma duração de quatro semanas, com sessões semanais de três horas.

O desenvolvimento contou com a participação de 12 estudantes do primeiro ano do Ensino Médio, com idades entre 14 e 15 anos e quivalência de gênero, oriundos de uma escola pública urbana localizada na região sudeste do estado de São Paulo, selecionados pela disponibilidade e pelo interesse em atividades extracurriculares focadas em inovação e sustentabilidade.Além disso, possuíam acesso básico a dispositivos eletrônicos, o que facilitou a compreensão e o engajamento nas atividades.

## 4. Resultados e Análises

Pela especificidade de composição, métodos de produção e de usabilidades, os resultados são apresentados e analisados por dispositivo produzido.

*3.1. Powerbank*

Um *powerbank* é uma bateria externa utilizada para carregar celulares em situações em que não há disponíveis fontes usuais (como tomadas elétricas). Para a sua montagem, foram reaproveitadas baterias cilíndricas de lítio Li-íon 18650 recarregáveis, provenientes de *laptops* em desuso. cada qual com voltagem nominal de 3,7 V e capaz de manter corrente de 2,2 A por até uma hora. A Figura 1 ilustra os estudantes em colaboração no espaço *maker* da escola para preparar e testar as baterias para a confecção do *powerbank*.

Os dispositivos comerciais são usualmente carregados a partir de uma tensão nominal (de saída) que varia de 5 a 9 V. Para alcançá-la no dispositivo construído, foi necessário conectar as baterias com capacidade de 2,2 Ah em série e em paralelo. Esse procedimento também faz com que o dispositivo tenha uma capacidade ampère-hora[4] (Ah) maior.

---

[4] O Ah indica a quantidade de energia transferida por uma corrente estável de 1 ampère (A) durante o período de 1 hora (h).

**Figura 1.** Preparo e teste das baterias de lítio retiradas de *laptops* fora de uso para a construção do *powerbank*.

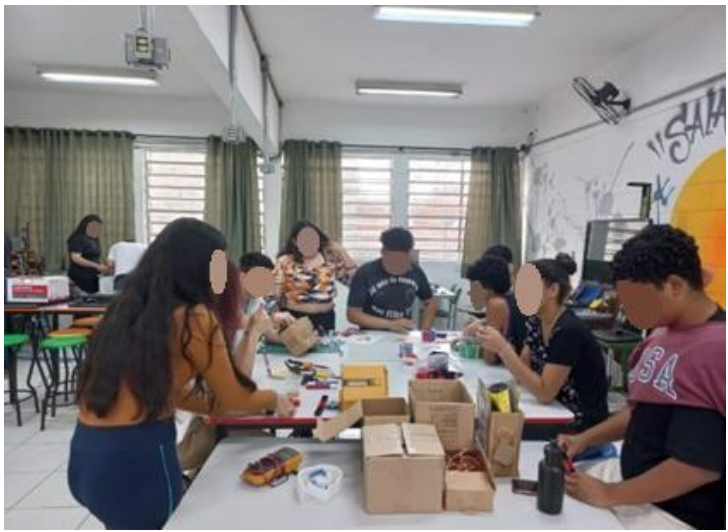

Fonte: Arquivo dos autores (2025).

Quando duas ou mais baterias são *conectadas em série,* sua capacidade é mantida constante e sua diferença de potencial é aumentada. Essa conexão é feita quando o terminal positivo de uma bateria é conectado ao terminal negativo de outra, sucessivamente, até que todas as baterias estejam conectadas. Se conectarmos, por exemplo, 3 baterias de 3,7 V e 2,2 Ah em série, teremos uma bateria equivalente com uma voltagem maior, de aproximadamente 11 V, e a mesma capacidade de 2,2 Ah.

Se as baterias forem *conectadas em paralelo*, a voltagem é mantida constante e a sua capacidade é aumentada. A conexão, neste caso, é estabelecida quando o terminal positivo de uma bateria é conectado ao terminal positivo de outra. O mesmo é feito com os terminais negativos. Nesse caso, se 3 baterias de 3,7 V e 2,2 Ah forem conectadas em paralelo, tem-se uma equivalente com tensão de 3,7 V e uma capacidade de 6,6 Ah.

Para a montagem do *powerbank*, foram utilizadas 9 baterias cilíndricas iguais de lítio Li-íon 18650 de 3,7 V e 2,2 Ah. Conectaram-se 3 baterias em paralelo formando 3 conjuntos de baterias de 3,7 V e 6,6 Ah e, em seguida, esses três conjuntos foram conectados em série para obtenção de uma bateria de aproximadamente 11 V e 6,6 Ah, conforme mostrado na Figura 2. Isto é, com essas conexões, é possível aumentar a tensão e a capacidade do *powerbank*, conforme a necessidade.

**Figura 2.** Foto das 9 baterias de lítio Li-íon 18650 de 3,7 V e 2,2 Ah conectadas em série e em paralelo para a confecção do *powerbank*, formando uma bateria equivalente de 11 V e 6,6 Ah. As conexões entre as baterias foram feitas com pequenas placas metálicas utilizando uma máquina de solda de ponto.

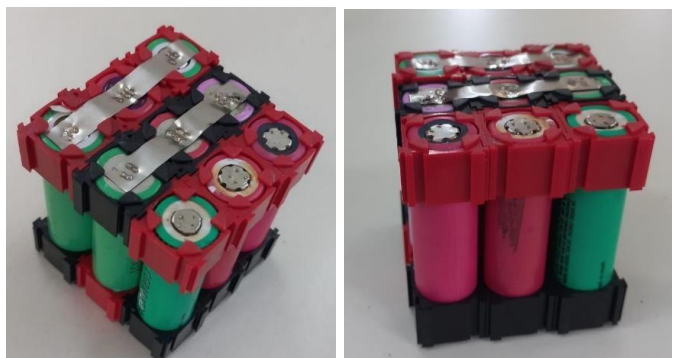

Fonte: Arquivo dos autores (2025).

Para que o conjunto de baterias funcione de maneira adequada, é necessário conectá-lo a uma placa BMS[5]. Esse dispositivo eletrônico é útil para protegê-las de curto-circuito, sobrecarga e sobretensão, permitindo que a tensão total do conjunto de baterias ligadas em série e em paralelo, fornecida por um carregador simples, chegue adequadamente às células. A placa utilizada suporta no máximo 20 A.

As conexões feitas na placa BMS seguem o valor nominal de tensão dos três conjuntos de baterias. O ponto de referência da placa, dada por 0 V, deve ser conectado no eletrodo negativo (-) do conjunto de baterias e o máximo de tensão da placa, dado por 12 V, é conectado no eletrodo positivo (+) das baterias. O eletrodo positivo do primeiro conjunto de baterias, com tensão nominal de 3,7 V, é conectado no terminal de 4,2 V da placa e o eletrodo positivo de dois conjuntos de baterias, com tensão de 7,4 V, no terminal de 8,4 V da placa BMS, conforme ilustrado na Figura 3.

**Figura 3.** Esquema ilustrativo mostrando as conexões entre as 9 baterias cilíndricas de lítio Li-íon 18650 de 3,7 V e 2,2 Ah, a placa BMS de 20 A, o módulo regulador de tensão DC DC *Step Down* com tensão de alimentação entre 6 e 32 V e tensão de saída de 5 V para carregamento rápido com saída USB, o medidor de porcentagem de carga XW228DKFR4 e o interruptor, para compor o *powerbank* construído nesse trabalho.

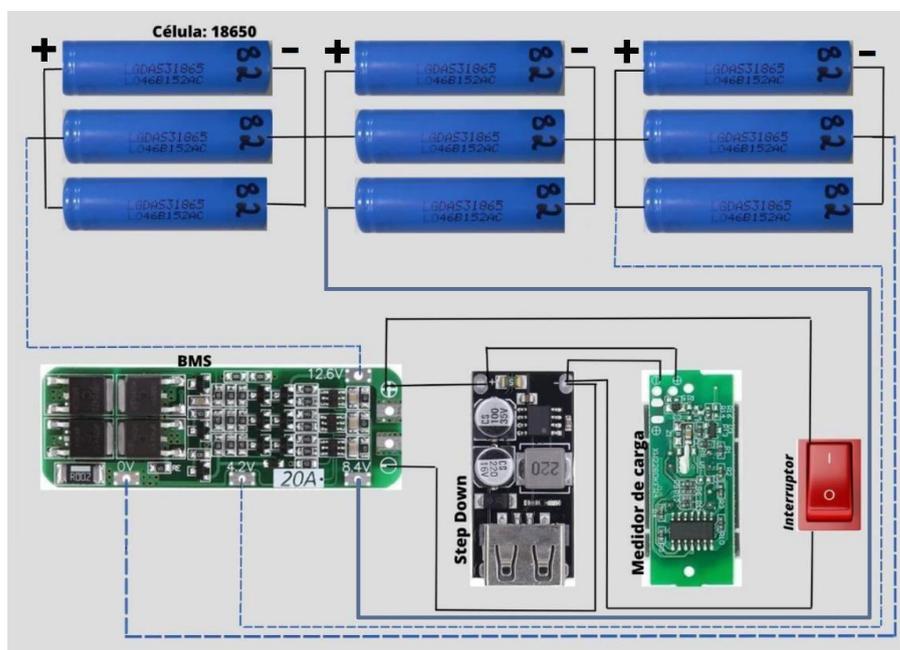

Fonte: Elaboração própria (2025).

Para utilizar essa montagem como um carregador de celulares, *smartphones* e *tablets*, é necessário reduzir sua tensão de saída de 11 para 5 V. Isso foi feito com um abaixador de tensão, também conhecido como *step down*, que consiste em um conversor DC-DC utilizado para diminuir a tensão de entrada de 11 V das baterias do nosso sistema para uma tensão de saída de 5 V para o carregamento dos dispositivos mencionados. Esse abaixamento de tensão é realizado às custas do aumento de corrente.

Para monitorar a carga total do *powerbank*, conectou-se um medidor de porcentagem de carga no *step down*. O dispositivo foi finalizado com a instalação de um interruptor liga-desliga conectado no *step down* e na placa BMS, como ilustrado na figura 3. Toda a montagem foi compactada e acondicionada em uma caixa feita em uma impressora 3D. Uma foto de dois dispositivos prontos é apresentada na Figura 4.

---

[5] BMS é um controlador de carga, isto é, de fluxo energético em baterias.

**Figura 4.** *Powerbanks* finalizados, prontos para serem utilizados para o carregamento de dispositivos eletrônicos, como celulares, smartphones e tablets. Para mostrar o seu funcionamento uma lâmpada foi conectada em sua saída USB.

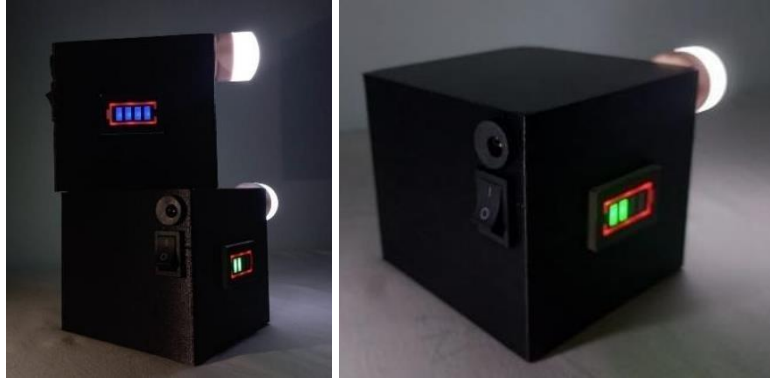

Fonte: Arquivo dos autores (2025).

Para carregar o *powerbank* construído numa tomada, totalmente descarregado, são necessárias de 3 a 4 horas. O tempo de carga de *smartphones* mais modernos no *powerbank* construído é de aproximadamente 3 horas. Para dispositivos mais antigos, a carga é realizada em tempo menor, uma vez que a capacidade das baterias, em Ah, também são relativamente menores. O *powerbank* construído pode ser utilizado ininterruptamente para carregamentos por um tempo médio de 15 horas.

### 3.2. Dispositivo carregador acoplado a uma bicicleta

Alguns veículos possuem dispositivos que aproveitam o seu próprio movimento para o carregamento das baterias, de modo a manter componentes eletrônicos funcionando continuamente. Nessa toada, os estudantes exploraram o movimento da roda da bicicleta para carregar as baterias de seus próprios *smartphones*.

O primeiro protótipo do dispositivo carregador foi acoplado a uma bicicleta em uma montagem estacionária, com a sua roda traseira suspensa e presa a um cavalete de madeira. Para a fonte de alimentação do dispositivo carregador nesta configuração, foi utilizado um dínamo de motor de skate elétrico (*hoverboard*) fixado a outro suporte de madeira para ficar em contato com a roda traseira da bicicleta, como mostrado na Figura 5.

**Figura 5.** Bicicleta apoiada em um cavalete de madeira, com o pneu em contato com o motor de um skate elétrico (*hoverboard*), para converter a energia mecânica da bicicleta em energia elétrica.

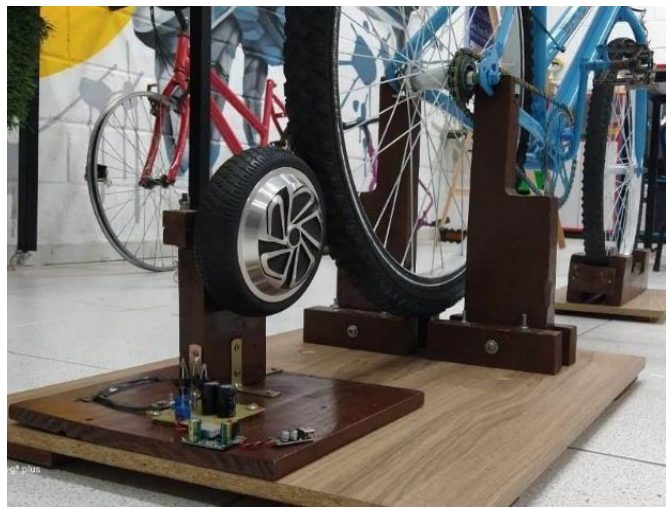

Fonte: Arquivo dos autores (2025).

O dínamo funciona como um gerador elétrico trifásico capaz de converter a energia mecânica proveniente da rotação do pneu da bicicleta em energia elétrica. Após gerada, a energia elétrica passa por um retificador de corrente para transformar a corrente alternada em contínua. Como a tensão gerada é alta, essa deve ser abaixada para que seja possível carregar celulares. Para isso, foram utilizados dois *step downs*. O primeiro, para abaixar a tensão gerada para 12 V; o segundo, como no caso do *powerbank*, para fornecer uma tensão de 5 V. O caminho da corrente no dispositivo é apresentado na Figura 6 e o circuito montado é apresentado na Figura 7.

**Figura 6.** Esquema do caminho percorrido pela corrente elétrica gerada no motor elétrico até à conexão USB para carregar dispositivos eletrônicos, mostrando o gerador trifásico, responsável por converter a energia mecânica em elétrica, seguido pelo retificador de corrente, para que uma corrente DC chegue até os *step downs* para abaixar a tensão de saída do dispositivo para 5 V.

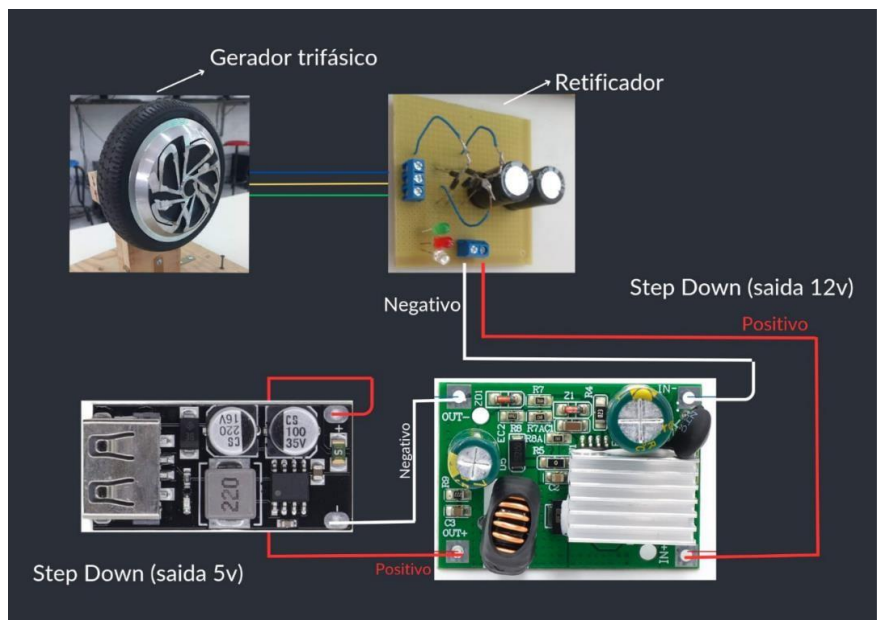

Fonte: Elaboração própria (2025).

**Figura 7.** Circuito do dispositivo carregador completo e montado para fornecer uma tensão de saída de 5 V para o carregamento de celulares.

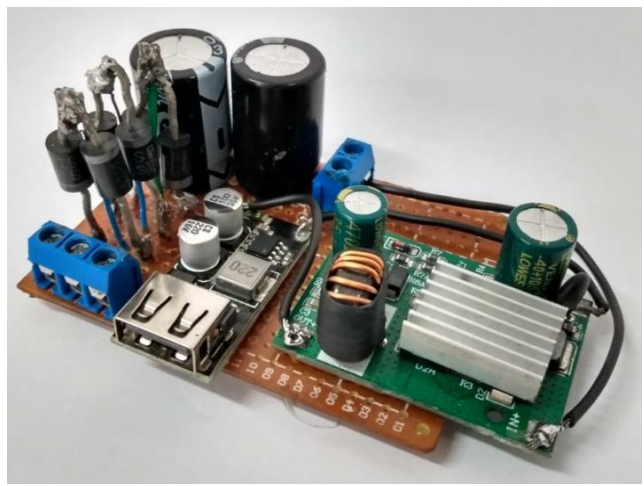

Com esse protótipo estacionário, os estudantes mostraram que é possívelcarregar dispositivos eletrônicos móveis e *powerbanks* pedalando-se a bicicleta. Além disso, analisaram a eficiência do sistema, verificando que é possível carregar os dispositivos eletrônicos em velocidades de passeio, entre 10 e 15 km/h. Com esseresultado, sentiram-se motivados para trabalhar em um sistema que pudesse ser implementado na bicicleta em movimento.

Para isso, foi necessário substituir o gerador trifásico acoplado à roda de *hoverboard* por um sistema menor, que pudesse ser adaptado e fixado à roda dianteira da bicicleta. Ele é constituído por um gerador de passos, reaproveitado o retificador de corrente e um *step down* de uma impressora 3D, fora de uso (Figura 8). Para testar o poder de carregamento do sistema com a energia elétrica gerada, foi utilizado um *powerbank*. Os resultados indicaram que o sistema tem potencial para realizar o carregamento dos dispositivos eletrônicos nas faixas de velocidades analisadas.

**Figura 8.** Visão lateral, à esquerda, e superior, à direita, do dispositivo carregador acoplado à roda dianteira da bicicleta para ser utilizado com a mesma em movimento.

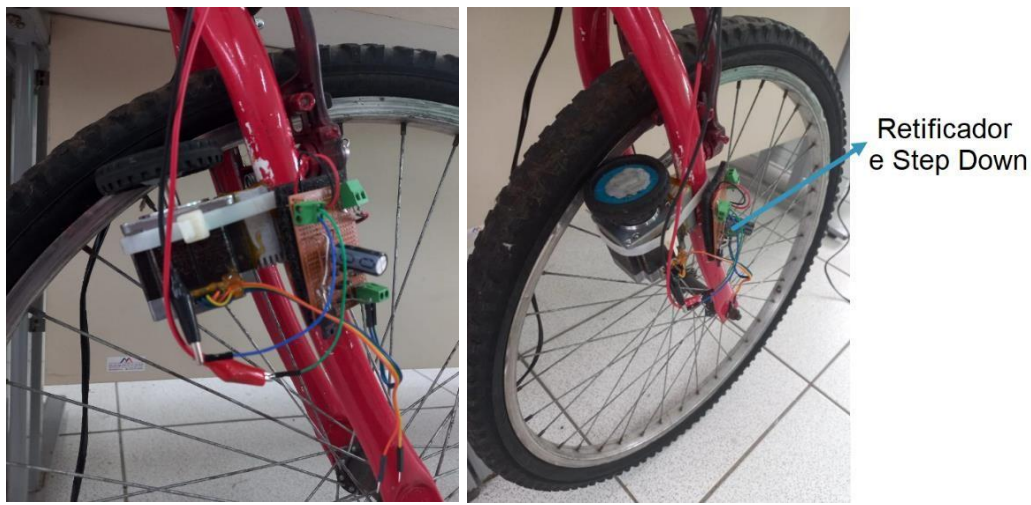

*3.3. Promovendo a prática de exercícios físicos no ambiente escolar e na comunidade*

Os estudantes realizaram, no ambiente escolar, excelente trabalho de divulgação acerca de compreensões relativas à energia limpa e sustentável e à importância de praticar atividades físicas. Para traduzir visualmente a proposta, criaram logotipo de divulgação com o *slogan* "Pedala Energia" (Figura 9). Ele é descrito por um ciclista no formato de raio, para remeter à conversão de energia mecânica em elétrica. A ideia de utilizar energia limpa de maneira sustentável aparece naturalmente com a presença de folhas verdes. Os polos positivo e negativo, envolvendo a bicicleta, deixam evidente o propósito do dispositivo: carregar baterias de aparelhos eletrônicos por meio das pedaladas. Isto é, a partir de uma atividade física saudável e não poluente, é possível gerar energia limpa para finalidades como recarregar a bateria de *smartphones* e *tablets*.

**Figura 9.** Logotipo criado pelos estudantes da escola para traduzir visualmente a proposta do projeto.

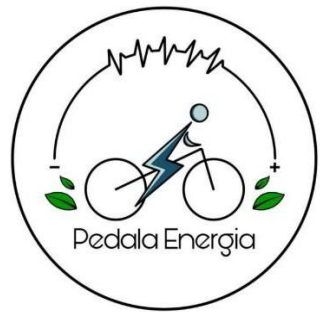

Fonte: Elaboração própria (2025).

Diante do sucesso de divulgação no ambiente escolar, os estudantes se mobilizaram para atingir o público externo, criando uma conta no Instagram com informações acerca do projeto e de sua importância para a geração deenergia limpa em uma proposta saudável e sustentável. Eles apresentaram dados do IBGE dedicados ao combate do sedentarismo, às suas causas, às doenças interligadas e à sua prevenção, remetendo ao Dia Nacional de Combate ao Sedentarismo, estabelecido em 10 de março (Brasil, 2023), conforme Figura 10.

**Figura 10.** Layout da página do Instagram para divulgação do projeto Pedala Energia e a importância de combater o sedentarismo.

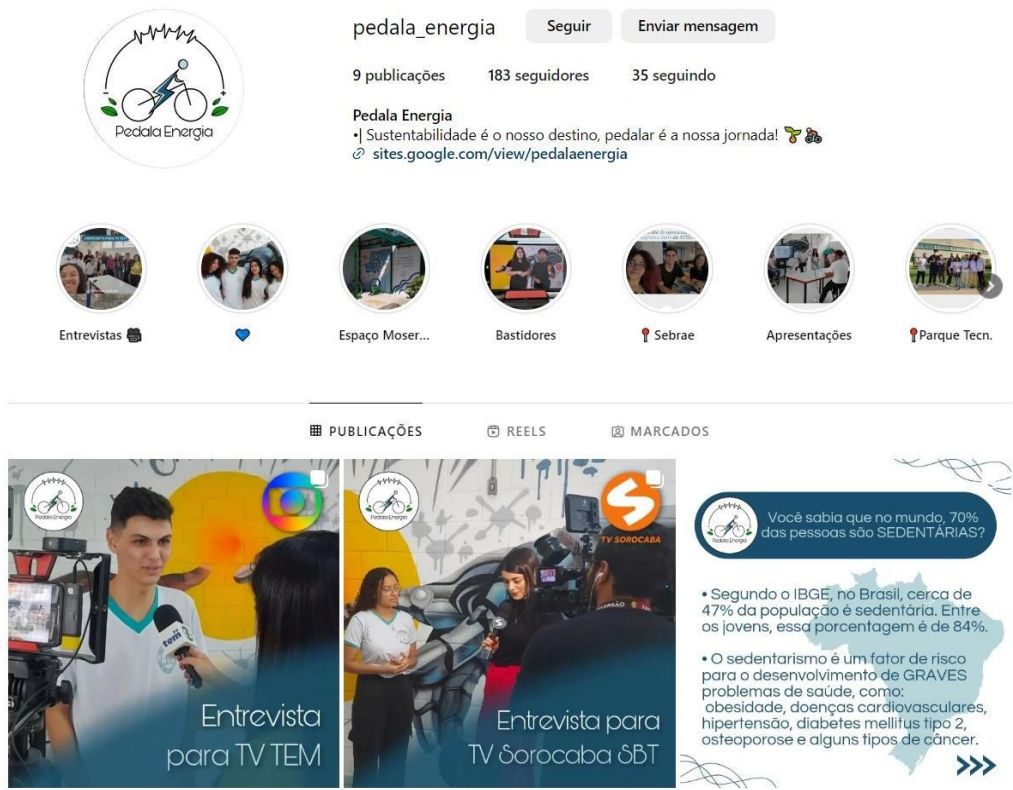

Fonte: Instagram. Disponível em: https://www.instagram.com/pedalaenergia/?igsh=bmY0eHNzdDhtNGYy. Acesso em 10 jan. 2025.

## 5. Considerações Finais

Um dos principais objetivos da pesquisa tecnológica descrita neste artigo foi a geração de energia limpa e sustentável a partir do reaproveitamento de materiais eletrônicos descartados, explorando a proposta de metareciclagem para a montagem de dois dispositivoscarregadores de *smartphones* e *tablets*. Ela permitiu que os estudantes envolvidos desenvolvessem conhecimentos teóricos do currículo de Física, considerando conceitos como carga e correntes elétricas contínua e alternada, campos elétricos e magnéticos, diferença de potencial, potência elétrica e efeito Joule, dispositivos eletrônicos, sistemas de transmissão, entre outros, bem como competências e habilidades práticas, como na exploração de materiais, a associação dos dispositivos eletrônicos, construção e caracterização de circuitos eletrônicos, a utilização de instrumentos de medição e a elaboração do próprio *design* do dispositivo, contextualizando efetivamente a cultura maker (Paula; Oliveira; Martins, 2019). Adicionalmente, mostraram possível incentivar a prática de exercícios físicos, principalmente entre os jovens estudantes, para combater o sedentarismo.

Cientificamente, foi de grande relevância tratar da necessidade de investir no currículo de disciplinas científicas visando implementar educação e cultura científicas na escola. Analisar e produzir argumentos científicos são processos importantes na produção de ciência, pois se baseiam na compreensão conceitual e não na memorização mecânica de fenômenos, definições e modelos matemáticos. O conhecimento acerca de tecnologias a partir da proposta de metareciclagem motivou os estudantes a estudarem conceitos e metodologias científicas como argumentação, fundamentação, alfabetização científica e comunicação e divulgação.

O desenvolvimento dessa proposta e das respectivas competências e habilidades resultaram em excelentes desempenhos dos estudantes em argumentações ao longo das aulas, competições e feiras científicas. Em 2022, foram campeões no projeto ECO TEAM *for a better world* idealizado pelo Centro Cultural Brasil – Estados Unidos (CCBEU) de Sorocaba. Ainda naquele ano, obtiveram a 2º posição na Feira STEM Brasil 2022 na modalidade Mobilidade Urbana e Transporte; em 2023, foram semifinalistas no programa global de cidadania corporativa promovido pela Samsung, intitulado *Solve for Tomorrow* Brasil, posicionando-se entre os 20 melhores trabalhos.

O reconhecimento adquirido com conquistas associadas a propostas viáveis, inteligentes e colaborativas entre a escola e a comunidade atesta a potencialidade de estudantes de escolas públicas no desenvolvimento científico e crítico-transformativo dos respectivos ambientes.. Em síntese, considera-se satisfatória a integração entre ciência e escola pública da EB para o desenvolvimento de tecnologias sustentáveis em que os estudantes desenvolveram autonomia e protagonismo.

**Referências**

Aguiar, E. S., Ribeiro, M. M., Viana, J. H., & Pontes, A. N. (2021). Panorama da disposição de resíduos sólidos urbanos e sua relação com os impactos socioambientais em estados da Amazônia brasileira. *Revista Brasileira de Gestão Urbana*, 13, e20190263. https://doi.org/10.1590/2175-3369.013.e20190263.

Baldé, C. P., Forti, V., Gray, V., Kuehr, R., & Stegmann, P. (2017). *The global e-waste monitor 2017: quantities, flows and resources*. Bonn: United Nations University (UNU), International Telecommunication Union (ITU) & International Solid Waste Association (ISWA). https://collections.unu.edu/eserv/UNU:6341/Global-E-waste_Monitor_2017__electronic_single_pages_.pdf.

Brasil. (2017). Ministério da Educação. *Base Nacional Comum Curricular (BNCC)*. Brasília: Ministério da Educação. https://www.gov.br/mec/pt-br/escola-em-tempo-integral/BNCC_EI_EF_110518_versaofinal.pdf.

Bunge, M. (1985). *Treatise on basic philosophy*. Part. II. Boston: D. Reidel.

Ferreira, M., Nogueira, D. X. P. ., Silva Filho, O. L. da, Costa, M. R. M., & Soares Neto, J. J.

(2022). A WebQuest como proposta de avaliação digital no contexto da aprendizagem significativa crítica em ciências para o ensino médio. *Pesquisa e Debate em Educação*, *12*(1), e35023. https://doi.org/10.34019/2237-9444.2022.v12.35023.

Ferreira, M., Silva Filho, O. L. da, Portugal, K. O., Bottechia, J. A. de A., Lima, M. B., Costa, M. R. M., Ferreira, D. M. G., & Oliver, N. A. D. (2022). Formação continuada de professores de Ciências em caráter investigativo, interdisciplinar e com mediação por tecnologias digitais: reflexões acerca do curso ciência é 10 na universidade de Brasília. *Revista Brasileira De Pós-Graduação*, *18*(39), 1–39. https://doi.org/10.21713/rbpg.v18i39.1971.

Ferreira, M., Silva Filho, O. L., Portugal, K. O., Bottechia, J. A. A., Lima, M. B., Costa, M. R. M., Ferreira, D. M. G., & Oliver, N. A. D. (2022). Formação continuada de professores de Ciências em caráter investigativo, interdisciplinar e com mediação por tecnologias digitais. *Revista Brasileira da Pós-Graduação*, 18(39), 1–39. https://doi.org/10.21713/rbpg.v18i39.1971.

Franz, N. M., & Silva, C. L. (2021). Resíduos de Equipamentos Elétricos e Eletrônicos (REEE): desafio global e contemporâneo às cadeias produtivas e ao ambiente urbano. *Gestão & Produção,* 29, Article e6621. https://doi.org/10.1590/1806-9649-2022v29e6621.

Freitas Junior, V., & Sousa, V. M. De. (2018). *Guia para escrita de artigos científicos: uma perspectiva da pesquisa tecnológica*. Sombrio: Instituto Federal Catarinense. https://redes.sombrio.ifc.edu.br/wp-content/blogs.dir/9/files/sites/84/2023/02/Guia-de-artigos-cientificos.pdf.

Gulis, G., Silva Filho, O. L., Ferreira, M., Andrade, V. C., & Costa, M. R. M. (2021). Ensino Interdisciplinar da Fotossíntese: Interfaces entre a Aprendizagem Significativa Crítica e as Comunidades de Investigação. *Experiências em Ensino de Ciências*, 16(3), 89–116. https://fisica.ufmt.br/eenciojs/index.php/eenci/article/view/974/852.

Halverson, E. R, & Sheridan, K. M. (2014). The Maker Movement in Education. *Havard Educational Review,* 84(4), 495–504. https://doi.org/10.17763/haer.84.4.34j1g68140382063.

https://ewastemonitor.info/wp-content/uploads/2020/11/GEM_2020_def_july1_low.pdf.

Ilankoon, I. M. S. K., Ghorbani, Y., Chong, M. N., Herath, G., Moyo, T., & Petersen, J. (2018). E-waste in the international context – A review of trade flows, regulations, hazards, waste management strategies and technologies for value recovery. *Waste Management*, 82, 258-275. https://doi.org/10.1016/j.wasman.2018.10.018.

Lévy, P. (2003). *A inteligência coletiva: por uma antropologia do ciberespaço*. São Paulo: Loyola.

Lévy, P. (2010). *Cibercultura*. São Paulo: Editora 34.

Lu, C., Zhang, L., Zhong, Y., Ren, W., Tobias, M., Mu, Z., Ma, Z., Geng, Y., & Xue, B. (2014). An overview of e-waste management in China. *Journal of MaterialCycles and Waste Management,* 17(1), 1-12. https://link.springer.com/article/10.1007/s10163-014-0256-8.

Mill, D. R. S., Oliveira, A. A., Ferreira, M. (2022). Jornadas formativas em tempos de pandemia: aportes para pensar atividades assíncronas. *Revista da FAEEBA: Educação e Contemporaneidade.* 31(65), 201-224. https://doi.org/10.21879/faeeba2358-0194.2022.v31.n65.p201-224.

Neto, J. S. (2022). *Processos educomunicativos na metareciclagem: formação de professores*

das fábricas de cultura 4.0 de São Paulo*. [Trabalho de Conclusão de Curso, Universidade de São Paulo]. USP Repository. https://bdta.abcd.usp.br/item/003140578.

Oliveira, A. A., Mill, D., & Ferreira, M. (2023). Jornadas formativas híbridas, invertidas e (signific)ativas no ensino superior: aportes para pensar atividades síncronas. *Revista Eletrônica de Educação*, 17, e5735107, 1-24. http://dx.doi.org/10.14244/198271995735.

Orti, V., Baldé, C. P., Kuehr, R., Bel, & G. (2020). *The global e-waste monitor 2020: Quantities, flows, and the circular economy potential*. United Nations University (UNU)/United Nations Institute for Training and Research (UNITAR) – co-hosted SCYCLE Programme, International Telecommunication Union (ITU) & International Solid Waste Association (ISWA), Bonn/Geneva/Rotterdam.

Paula, B. B., Oliveira, T., & Martins, C. B. (2019). Análise do Uso da Cultura Makerem Contextos Educacionais: Revisão Sistemática da Literatura. *Revista Novas Tecnologias na Educação*, 17(3), 447–457. https://doi.org/10.22456/1679-1916.99528.

Silva Filho, O. L., & Ferreira, M. (2018). Teorias da Aprendizagem e da Educação como Referenciais em Práticas de Ensino: Ausubel e Lipman. *Revista do Professor de Física,* 2(2), 104–125. https://doi.org/10.26512/rpf.v2i2.12315.

Sommerman, A. (2015). Objeto, Método e Finalidade de Interdisciplinaridade. *In*: A. Philippi Jr & V. Fernandes (Eds.). *Práticas de interdisciplinaridade no ensino e pesquisa* (pp. 165–212). Barueri: Manole.

UN DESA. (2021). *The Sustainable Development Goals Report 2021*.New York, USA: UN DESA. https://unstats.un.org/sdgs/report/2021/The-Sustainable-Development-Goals-Report-2021.pdf.